\documentclass[sigconf]{acmart}
\AtBeginDocument{%
  }

\setcopyright{acmlicensed}
\copyrightyear{2018}
\acmYear{2018}
\acmDOI{XXXXXXX.XXXXXXX}
\acmConference[Conference acronym 'XX]{Make sure to enter the correct
  conference title from your rights confirmation email}{June 03--05,
  2018}{Woodstock, NY}
\acmISBN{978-1-4503-XXXX-X/2018/06}

\usepackage{subfiles}

\newif\ifusecolor
\usecolorfalse   

\PassOptionsToPackage{table,xcdraw}{xcolor}
\usepackage{colortbl}
\usepackage{cleveref}
\usepackage{todonotes}
\usepackage{booktabs}
\usepackage{xcolor}
\usepackage{colortbl} 

\newcommand{\revisionTable}[1]{%
  {\color{blue}#1}%
}
\newcommand*\iftodonotes{\if@todonotes@disabled\expandafter\@secondoftwo\else\expandafter\@firstoftwo\fi}  

\usepackage{graphicx}
\usepackage{enumitem}

\usepackage{multirow}
\usepackage{booktabs}
\usepackage{array}

\copyrightyear{2026}
\acmYear{2026}
\setcopyright{cc}
\setcctype{by}
\acmConference[CHI '26]{Proceedings of the 2026 CHI Conference on Human Factors in Computing Systems}{April 13--17, 2026}{Barcelona, Spain}
\acmBooktitle{Proceedings of the 2026 CHI Conference on Human Factors in Computing Systems (CHI '26), April 13--17, 2026, Barcelona, Spain}
\acmPrice{}
\acmDOI{10.1145/3772318.3791639}
\acmISBN{979-8-4007-2278-3/2026/04}
\begin{document}

\title{Beyond Content Exposure: Systemic Factors Driving Moderators' Mental Health Crisis in Africa}

\author{Nuredin Ali Abdelkadir}
\affiliation{%
  \institution{Computer Science and Engineering}
  \institution{University of Minnesota}
  \institution{The Distributed AI Research Institute}
  \state{Minnesota}
  \country{USA}}
\email{ali00530@umn.edu}

\author{Tianling Yang}
\affiliation{%
  \institution{Weizenbaum Institute for the Networked Society}
  \institution{Technische Universität Berlin}
  \city{Berlin}
  \country{Germany}}
\email{tianling.yang@tu-berlin.de}

\author{Shivani Kapania}
\affiliation{%
  \institution{Carnegie Mellon University}
  \city{Pittsburgh}
  \country{USA}}
\email{kapania@cmu.edu}

\author{Kauna Ibrahim Malgwi}
\affiliation{%
  \institution{Pan African Christian University}
  \country{Kenya}}
\email{kauna.ibrahim@students.pacuniversity.ac.ke}

\author{Fasica Berhane Gebrekidan}
\affiliation{%
  \institution{ Independent Researcher}
  \country{Kenya}}
\email{fasikaberhane@gmail.com}

\author{Adio-Adet Dinika}
\affiliation{%
  \institution{The Distributed AI Research Institute}
  \country{Germany}
  }
\email{adio@dair-institute.org}

\author{Elaine O. Nsoesie}
\affiliation{%
    \institution{School of Public Health}
    \institution{Boston University}
  \institution{The Distributed AI Research Institute}
  %
  \city{Massachusetts}
  \country{United States}
  }
\email{onelaine@bu.edu}

\author{Milagros Miceli}
\affiliation{%
  \institution{The Distributed AI Research Institute}
  \institution{Technische Universität Berlin}
  \city{Berlin}
  \country{Germany}
  }
\email{mila@dair-institute.org}

\author{Stevie Chancellor}
\affiliation{%
  \institution{Computer Science and Engineering}
  \institution{University of Minnesota}
  \state{Minnesota}
  \country{USA}}
\email{steviec@umn.edu}

\renewcommand{\shortauthors}{Abdelkadir et al.}
\newcommand{\revision}[1]{\textcolor{black}{#1}}

\begin{abstract}

Content moderators review disturbing content to protect social media users, often at significant cost to their mental health. Recent reports document the mental health conditions of African moderators as notably problematic. Beyond the content itself, what factors contribute to the deteriorating mental health of these workers? We surveyed 134 moderators across Africa to understand their mental health and interviewed 15 moderators to contextualize their experiences. We found that African moderators suffer from high psychological distress and lower well-being compared to moderators in other areas. Former moderators showed significantly higher distress levels, demonstrating long-term impact that extends beyond their moderation work. Our interviews showed that systemic and structural labor conditions contribute to moderators’ severe psychological distress and diminished mental well-being. Corporate wellness programs promoted by platforms were found ineffective and inadequate. We discuss how this requires holistic attention and structural solutions by all involved parties to improve moderators' mental health.

\end{abstract}

\ccsdesc[500]{Human-centered computing~Empirical studies in HCI}
\ccsdesc[500]{Human-centered computing~Social media}
\keywords{Social media, content moderation, mental health, critical HCI, data work, psychological distress, mental well-being, outsourcing, content exposure}



\maketitle

\label{sec:introduction}
\section{INTRODUCTION}


Social media platforms depend on an invisible workforce of content moderators who shield billions of users from harmful online content \cite{roberts2019behind, gillespie_custodians_2018, gray2019ghost, dark_ghost_2023, steiger2021psychological}. Tens of thousands of moderators are employed by third-party companies to review user-generated content ranging from hate speech to unimaginably extreme images \cite{barrett2020moderates}. Although platforms like Facebook, TikTok, and YouTube rely on algorithmic filtering, human moderators remain essential for contextual judgment calls, cultural sensitivity, and handling edge cases that algorithms cannot reliably process \cite{gillespie_custodians_2018}. This work is typically outsourced to Business Process Outsourcing (BPO) companies, and this digital labor comes at a profound human cost, significantly affecting moderators' mental health~\cite{spence_psychological_2023,spence2025content, steiger2021psychological}.

In recent years, the deteriorating mental health of African content moderators has gained international attention \cite{bhalla_feature-mental_2023, kleinman_i_2024, booth_ptsd_2024, b_investigative_journalists_metas_2025}. 
In December 2024, The Guardian reported on the outcomes of a medical report in Kenya: ``more than 140 Facebook content moderators have been diagnosed with severe post-traumatic stress disorder caused by exposure to graphic social media content including murders, suicides, child sexual abuse and terrorism." \cite{booth_more_2024}. 
In follow-up personal accounts, African moderators have shared experiences of how this work has fundamentally altered their mental health, relationships, and worldview \cite{gebrekidan_content_2024, Botlhokwa_content_2024}. Africa, in particular, has become a hub for content moderation outsourcing, with countries like Kenya, Ghana, and Tunisia hosting major BPO operations serving primarily Western platforms and users \cite{berthelot_tiktoks_2023, wilmont_metas_2025, elswah2024moderating}. 

Graphic content exposure is often cited as one of the primary reasons for these adverse mental health impacts~\cite{booth_more_2024, kleinman_i_2024, booth_ptsd_2024, kimeu_work_2024}. This area of interest has spun up research in HCI and adjacent Computer Science fields to build better tools for moderators engaging with this content, including hiding and blurring tools~\cite{das_fast_2020, dang2018but, karunakaran2019testing, gillespie_custodians_2018} and AI-assisted moderation to identify `not safe for work'~\cite{leu2024auditing,lee2024exploring} and severe health content~\cite{chancellor_norms_2018,chancellor_quantifying_2016,coppersmith_quantifying_2015}. However, while the traumatic nature of content exposure is well-documented, many other factors can influence moderators' work and well-being~\cite{steiger2021psychological, newton_secret_2019, wilmont_metas_2025}. Outsourcing companies often leave workers in precarious working situations, adding to their mental str\revision{ain} in navigating online content and harsh labor arrangements ~\cite{muldoon_feeding_2024, gebrekidan_content_2024}. 

What structural and workplace factors contribute to moderators' psychological distress, in addition to graphic content exposure? What is the comparative mental health status of African moderators relative to their counterparts in other regions? How do mental health outcomes vary across platforms (such as Meta and TikTok)? Most importantly, how do these conditions persist long after moderators leave the profession, and what interventions are needed by involved parties -- including BPOs, platforms, and governments -- to protect these workers' mental health? These questions are urgent given the geographic concentration of content moderation work in the Global South, where regulatory protections may be weaker and economic vulnerabilities higher. 

In this work, we address these questions through a mixed-methods survey and interview study of the psychological distress of Meta and TikTok content moderators in Africa. We conducted a comprehensive survey study ($n = 134$) to measure psychological distress and mental well-being, followed by in-depth semi-structured interviews ($n = 15$) to contextualize and better understand our quantitative findings. We found that over half ($55\%$) of moderators experience severe or moderate to severe psychological distress. The vast majority ($78\%$) of Meta moderators suffer from severe or moderate-to-severe psychological distress, while about half ($47\%$) of TikTok moderators fall into the same category. Former moderators exhibit significantly higher psychological distress than their current counterparts ($t=-2.145$, $p=0.0352$), demonstrating the persistent, long-term trauma that affects them months and years after leaving their positions. We found that African moderators experience higher psychological distress ($t=11.97$, $p<0.0001$) and lower mental well-being ($t=4.79$ $p<0.0001$) compared to moderators in other regions (i.e., Europe, Asia).

Our qualitative findings highlight how systemic and structural problems contribute to the deteriorating mental health conditions of moderators in Africa. Regardless of the platforms and BPOs moderators worked for, we found that precarious working conditions contribute to their declining mental well-being. These problems stem from extreme working conditions that encompass basic necessities for migrant workers (such as expired work permits), toxic work environments, failed corporate wellness programs, and poor compensation—all of which exacerbate moderators' mental health conditions and lead to long-term consequences. Moderators called for better design and implementation of technologies, improved mental health support, better working conditions, and fostering social recognition. We discuss the policy failures that sustain these extreme working conditions, resulting in African moderators experiencing higher psychological distress compared to other regions. We also examine the tensions researchers should consider when working with content moderators and argue for the need to move beyond design implications and solutions as a community to address these identified structural challenges. Our research contributes to the growing body of work on digital labor conditions while centering the voices and experiences of the content moderators themselves.

\label{sec:rw}
\section{BACKGROUND AND RELATED WORK}

\subsection{Commercial Content Moderation in Africa}

Many factors have shaped the rapid expansion of commercial content moderation in Africa. Some reports suggest the moderation industry expanded to the continent following the Arab Spring, as platforms sought to control social media movements and political unrest \cite{joseph2012social}.
The region's competitive labor market~\cite{perrigo_inside_2022, anwar2022digital}, and countries making digital outsourcing a pillar for economic growth~\cite{apofeed_kenya_2024} to address high unemployment~\cite{hoppner_africas_2025}, may have increased the content moderation industry. \revision{Other contributing factors include the continent's increasing youthful and educated workforce available during the current unemployment crisis, strong English-language skills (Kenya, Nigeria, where English is a national language), timezones convenient for business in America and Europe, and labor-cost arbitrage compared to Asian workers (Philippines)~\cite{the_economist_call_2025, hoppner_africas_2025}.} The BPO industry in Africa is expected to grow 14\% by 2028, nearly twice as fast as the global BPO growth rate of 8\% \cite{the_economist_call_2025}.

This industry's formalization began in early 2019 when Facebook contracted with Samasource (now Sama) to provide content moderation services for Sub-Saharan Africa markets ~\cite{perrigo_inside_2022}. Sama began hiring moderators across sub-Saharan countries and relocating them to its Kenya hub ~\cite{perrigo_inside_2022}. Soon, these companies expanded beyond moderating social media platforms. Kenyan moderators later served in safety screening for ChatGPT, labeling textual descriptions of sexual abuse, hate speech, and violent content from the darkest corners of the internet sent by OpenAI to its contracted outsourcing company, Sama ~\cite{perrigo_exclusive_2023}. In August 2023, TikTok began content moderation operations in Kenya ~\cite{france-presse_kenya_2023}.

Today, multiple outsourcing companies operate across the region to provide content moderation services. Platforms maintain stations across East, West, and North Africa to facilitate moderation work, employing hundreds to thousands of moderators. Teleperformance manages the moderation services for Meta products (Facebook and Instagram) and TikTok ~\cite{berthelot_tiktoks_2023, wilmont_metas_2025}. Teleperformance and Concentrix provide similar services for tech companies in North Africa ~\cite{elswah2024moderating}.

\subsection{Content Moderation Working Conditions}

For years, content moderation work occurred behind a veil of secrecy~\cite{roberts2019behind, newton_secret_2019}. In recent years, this ``ghost work''~\cite{gray2019ghost} has received attention through scholarly research and journalism documenting moderators' working conditions~\cite{roberts2019behind, roberts2016commercial, hern_facebook_2021, duffy_tiktok_2022, gilbert_facebook_2020}. 
Though automated content moderation is marketed widely by platforms, human moderators working under extreme conditions either take down posts or produce training data for automated moderation systems~\cite{roberts2019behind, newton_secret_2019}. Prior studies and reports document these conditions globally, including the Philippines, United States, India, and Estonia~\cite{savio2023social,bharucha2023content}. For example, Colombia investigated moderators' experience of alleged union-busting, traumatic working conditions, and low pay~\cite{mcintyre_traumatized_2022, perrigo_tiktoks_2022}. Similarly, moderators in different regions for non-English languages (Maghrebi Arabic Dialects, Kiswahili, Tamil, and Quechua) are often exploited, overworked, and underpaid \cite{elswah2025content}. \revision{Moderators'} conditions are \revision{created} through various mechanisms, such as Non-Disclosure Agreements, pressure to complete reviewing tasks in seconds, constant monitoring, ever-changing policies, and unrelenting pressure to meet quality standards~\cite{duffy_tiktok_2022, hern_facebook_2021, gilbert_facebook_2020}.

In the past two years, African content moderators have \revision{revealed their working experiences in BPOs for giant social media platforms} through the Data Workers' Inquiry (DWI) project \cite{DWI_2024}. These experiences highlight traumatic experiences for women, low payments, toxic work environments, and inadequate mental health \revision{support} \cite{gebrekidan_content_2024, Botlhokwa_content_2024}. \revision{M}oderators work eight to ten hours daily reviewing disturbing content. Moderators must make moderation decisions and cannot skip, even when content is too graphic, and sometimes work at twice the speed to ensure most videos are moderated or labeled \cite{hoppner_africas_2025, kgomo_i_2025}. Moderators describe working for these companies as inhumane, being treated as 'robots' with expectations to meet increasingly high metrics \cite{bradbury_factory_nodate}. To avert these extreme working conditions and hold BPOs and platforms accountable, \revision{over} 150 moderators came together to establish the African Content Moderators Union \cite{perrigo_150_2023}. Moderators are taking legal action against social media platforms and their outsourcing companies for harsh treatment and working conditions \cite{hall_meta_2025, perrigo_meta_2022}. Moderators continue to organize and speak at international venues, telling their stories about working conditions \cite{gebrekidan_content_2024, malgwi_content_2025, noauthor_8th_nodate, genderingaiconference_kauna_nodate}.

\subsection{Mental Health in Content Moderation}

Prior work has studied the effect of content moderation on moderators' mental health~\cite{steiger2021psychological, spence_psychological_2023, spence2024content}. Moderators face negative cognitive and emotional changes that can lead to behavioral changes, including avoidance, hypervigilance, and trauma responses~\cite{spence_psychological_2023}. They deal with intrusive thoughts, anxiety symptoms, and insomnia directly associated with the job, which may correspond to adjustment disorder~\cite{benjelloun2020psychological}.
Moderators' mental health experiences are comparable to first responders in other professions, such as emergency and social service workers~\cite{steiger2021psychological}. Moderators across different regions have clinical depression levels higher than law enforcement dealing with child sexual abuse material~\cite{spence2024content}.

Closest to our work are surveys of psychological well-being of moderators by Spence and coauthors.~\citet{spence2024content} surveyed moderators across North and South America, Asia, and primarily Europe \revision{revealed that} 
moderators with daily exposure to graphic content have higher psychological distress and lower well-being. In ~\citet{spence2025content}\revision{'s study,} 
\revision{60.6\%} of \revision{the surveyed moderators who are contracted to an Asian BPO} reported the wellness service provided by the BPO made them feel valued/heard. To address moderators' mental health experience, interventions have been suggested at different stages of the moderation process~\cite{steiger2021psychological}.~\citet{roberts2023we} found worker autonomy, peer-to-peer support, financial security, and respect within the larger company were fundamental to moderators' mental stamina. Social support, role validation, work-life balance, and individual therapy were also important individual and organizational coping mechanisms~\cite{spence2023content}.

In Africa, a psychiatric evaluation reported by the Guardian conducted on 140 Meta moderators showed severe psychological conditions~\cite{booth_more_2024}. Ex-Meta moderators were experiencing PTSD, depression, anxiety, and sometimes battling suicidal thoughts due to the job's effect~\cite{booth_ptsd_2024, kimeu_work_2024, kiplagat_ptsd_2024, hoppner_africas_2025}. They expressed continuous\revision{ly} deteriorating working conditions and being seen as moving objects with no care for their mental health, even after changing locations and working for different outsourcing companies~\cite{b_investigative_journalists_metas_2025}. The Data Worker's Inquiry project highlighted strenuous working conditions when moderating during a genocidal civil war~\cite{gebrekidan_content_2024}. Finally, \citet{malgwi_content_2025} outlined a proposal for better mental health conditions for African moderators that includes moderator-led group sessions.

This work expands our understanding of mental health conditions by outlining \revision{moderators'} mental health conditions and experiences, the systemic labor exploitation that \revision{damages} moderators' mental health, and the permanent and long-term consequences on moderators' psychological well-being\revision{. We focus} on the African content moderation industry\revision{, incorporating} views of moderators across different platforms, BPOs, and various markets/regions. Through a mixed-methods approach, we provide a holistic and scientific view of factors primarily \revision{driving} the worsening \revision{of moderators' }mental well-being during and after their moderation.

\label{sec:method}

\section{METHODS}

In this section, we outline methods for our mixed-methods study to understand moderators' mental health during and after employment. First, we overview our online survey \revision{with} 134 content moderators, then describe our online interviews with 15 moderators who completed the survey. The survey provided formal understanding of psychological distress and mental well-being levels of current and former content moderators \revision{across} different BPOs, factors contributing to their mental health, and suggestions for improving conditions. The in-depth interviews allowed us to better understand and contextualize factors affecting moderator mental health. \revision{Given that this was an online-only study,} the Institutional Review Board at the University of Minnesota, with study IDs STUDY00025681 and STUDY00025043, approved both the survey and interviews.

\subsection{Survey}

The survey incorporated self-reported assessment questionnaires aimed at understanding psychological distress and mental well-being of content moderators. The survey gathered participants' background information, such as whether they are current or former moderators, platforms they moderated, BPOs they work/worked for, country of focus, and language market. We used two clinical assessment questionnaires, the Clinical Outcomes in Routine Evaluation 10 (CORE-10)~\cite{barkham2013core} and Short Warwick-Edinburgh Mental Well-being Scale (SWEMWBS)~\cite{stewart2009internal}---designed to measure psychological distress and mental well-being, respectively. We finally asked moderators to describe main factors contributing to their mental health status, and their recommendations for addressing negative factors and improving their mental health. \revision{Appendix \ref{surevey_sample_questions_appendix} overviews the survey questions}. 

\subsubsection{Measures for Mental Well-Being}
We used two self-reported measures of moderators' psychological distress and mental well-being: the CORE-10 and SWEMWBS. Beyond their usage in general clinical assessment, psychologists in traumatic experiences have used these measures to study moderators' psychological and mental well-being conditions \cite{spence2024content, spence2025content}. Thus, they allow us to compare to general clinical and population baselines, as well as baselines of content moderators globally.

\textbf{CORE-10}: The Clinical Outcomes in Routine Evaluation 10 (CORE-10) \cite{barkham2013core} is a scale that measures psychological distress in mental health\footnote{\url{https://novopsych.com/assessments/outcome-monitoring/clinical-outcomes-in-routine-evaluation-10-core-10/}}. 
It assesses the experiences of those with depression, anxiety, PTSD, and other highly prevalent mental health disorders. This measure is a shortened version of the 34-item CORE-OM \cite{evans2002towards}. The survey asks participants experiences and feelings over the past week. Participants rate questions on a Likert Scale [0-4]. Scores range from 0 to 35, and higher scores indicate increased distress. The CORE-OM survey is an acceptable assessment tool in Kenya \cite{falkenstrom2018factor}, and CORE-10 is an acceptable reduced measure that compares in reliability to CORE-OM \revision{\cite{barkham2013core}}.    

\textbf{Short Warwick-Edinburgh Mental Well-being Scale}: 
This scale, which is often known as SWEMWBS \cite{stewart2009internal} is a clinical scale designed to understand and monitor positive mental well-being\footnote{\url{https://www.corc.uk.net/outcome-measures-guidance/directory-of-outcome-measures/short-warwick-edinburgh-mental-wellbeing-scale-swemwbs/}}. The measure contains seven questions about feelings and functioning of mental well-being, rated on a Likert Scale [1-5]. The measure has been adopted and validated in African countries (Tanzania and South Africa) \cite{smith2018validation, oyebode2023swahili}, helping ensure validity and reliability of our survey in the general African context. For SWEMWBS, scores range from 7 to 35; higher scores correlate with more positive well-being.

\subsubsection{Participant Recruitment}
\label{participant_recruitment}

We recruited participants by distributing flyers to online groups of African commercial content moderators on social media platforms. We worked with content moderators who are members of these communities to distribute flyers and facilitate recruitment. We distributed our survey to moderators in East, West, and North Africa; most BPOs and moderators are stationed in these three regions, with the majority in East Africa. We recruited participants who moderated for Meta and TikTok, spoke English, and were over 18. Most moderators in the region work for BPOs contracted to these two platforms. 

The survey took 12-15 minutes to complete on average. Before beginning, participants received a description of clinical questions with notice and warning that some questions could trigger prior mental health experiences from their moderation. A pilot was conducted with three moderators to ensure survey question clarity before wider deployment. We dropped participants who took less than 5 minutes to complete the survey for quality assurance (n=14 drops). There were 134 valid respondents: 44 current TikTok moderators, 34 former TikTok moderators, 49 former Meta moderators, and 7 other moderators. We did not recruit current Meta moderators due to the recent investigative report that disclosed their top-secret new site \cite{b_investigative_journalists_metas_2025}. These moderators may be closely scrutinized for involvement in our research. Survey participants resulted in moderators from 12 countries: Kenya, Ethiopia, Somalia, Nigeria, Uganda, South Africa, Tanzania, Rwanda, Burundi, Mali, Tunisia, and Morocco. Moderators received \$15 USD for completing the survey. Table \ref{tab:survey_demographics} summarizes participant demographics, and \revision{further details about survey participants, specifically former moderators, can be found in Appendix \ref{survey_participants_detail_appendix}.}

\begin{table}[htbp]
\centering
\begin{tabular}{@{}llc@{}}
\toprule
\textbf{Category} & \textbf{Group} & \textbf{Frequency (\%)} \\
                  &                & \textbf{N = 134} \\
\midrule
\multirow{3}{*}{\textbf{Gender}} 
                  & Female         & 55 (41.0\%) \\
                  & Male           & 78 (58.2\%) \\
                  & Non Binary     & 1 (0.7\%) \\
\cmidrule(lr){1-3}
\multirow{3}{*}{\textbf{Platform}} 
                  & Meta           & 49 (36.56\%) \\
                  & TikTok         & 78 (58.2\%) \\
                  & Other          & 7 (5.22\%) \\
\cmidrule(lr){1-3}
\multirow{4}{*}{\textbf{Experience}} 
                  & $<$1 year      & 10 (7.5\%) \\
                  & 1--2 years     & 59 (44.0\%) \\
                  & 2--3 years     & 45 (33.6\%) \\
                  & $>$3 years     & 20 (14.9\%) \\
\cmidrule(lr){1-3}
\multirow{2}{*}{\textbf{Moderator Status}} 
                  & Current        & 51 (38.05\%) \\
                  & Former         & 83 (61.95\%) \\ 
\bottomrule
\end{tabular}
\caption{Demographic data of survey participants. Platform BPO providers: Meta (Sama Source or Teleperformance/Majorel), TikTok (Teleperformance/Majorel, Concentrix). We keep the market or country of the survey participants private to protect them from targeted retaliation by the BPOs or platforms.}
\label{tab:survey_demographics}
\end{table}

\subsubsection{Survey Analysis}

Next, we describe how we analyzed the survey data. We quantitatively measured the clinical scores of the psychological distress and mental well-being assessments. The CORE-10 for each moderator is generated by summing their rating of each question\footnote{\url{https://novopsych.com/assessments/outcome-monitoring/clinical-outcomes-in-routine-evaluation-10-core-10/}}. The SWEMWBS scores are generated by summing the individual ratings and converting them into the appropriate scale as recommended by the creators of the assessment\footnote{\url{https://warwick.ac.uk/services/innovations/wemwbs/how/2025-02-res-area-ww-licenced}}. \revision{We compare African moderators psychological distress and mental wellbeing with their counterparts across other regions (i.e., Europe, Asia) from results generated by other studies. We then conduct a multivariate exploratory analysis using an Ordinary Least Squares model (OLS) on factors that potentially contribute to moderators' psychological distress. The confounding variables gathered in our survey include, worked duration (ordinal) where <1 years=1, 1-2=2, 2-3=3, >3=4); frequency of exposure (ordinal): where rare exposure=1, few times a month=2, few times a week=3, multiple times a week=4, once a day=5, few times a day=6, and multiple times a day=7; months since stopped working as a moderator, a continuous variable (i.e., current moderators= 0 months, etc), categorical variables such as, platform (Meta, TikTok, Other); BPO worked for (Teleperformance/Majorel, Sama Source, Concentrix, Other); gender (female, male, non-binary); Binary variables include: medications/drug use (yes=1, no=0), other current employment status for former moderators (yes=1, no=0); current work as a moderator (yes=1, no=0).\footnote{\revision{Ordinary Least Squares modeling was used, as the assumptions for the OLS model, such as normal residuals, linearity, and homoskedasticity, were met. We drop highly correlated variables using the variance inflation factor (VIF) -- we dropped BPO using this analysis.}}} 
Additionally, we provide demographic and descriptive statistics of the survey participants. 
Finally, we conducted an inductive thematic analysis \cite{kiger2020thematic} on the free-text and open-ended responses to understand the qualitative responses of the moderators to identify the major themes.

\subsection{Interviews}

To contextualize the survey findings and understand factors contributing to moderators' mental health, we conducted semi-structured interviews with 15 content moderators.

\subsubsection{Participant Recruitment} We recruited interview participants from survey respondents, sampling across countries/markets, platforms, BPOs, and working status (current or former). \revision{We selected participants to include diverse perspectives along these categories and based on the level of details of their qualitative responses in the survey.} 
Since interviews involved discussing moderators' mental health experiences, moderators could skip questions or stop if uncomfortable or triggered. \revision{For interviews and surveys, we provided mental health resource contacts (Emergency Medicine Kenya Foundation, National Emergency Hotline for Nigeria, and Emergency medical assistance service Casablanca for Morocco). We also provided World Health Organization resources/links for managing distress. Additionally, our team's clinical psychologist and former content moderator trained the first author (who conducted interviews) on less triggering interview techniques.} Interviews lasted 50-90 minutes, averaging 59 minutes. Moderators received \$50 USD. Table \ref{participants} shows interviewed moderators, their market, and platforms.

\begin{table}[t!]

\centering
\small
    {

  \begin{tabular}{lllc}
    \toprule
        Participant \# &\textbf{Pseudonym} & \textbf{Country Market} & \textbf{Platform} \\
    \midrule
    P1 & Shuman & Kenya &  CT \\ \hline
    P2 & Tanisha & Nigeria & CT \\ \hline
    P3 & Badru & South Africa, Ethiopia & CT \\ \hline
    P4 &  Zeyneb & Somali, Ethiopia & FT \\ \hline    
    P5 & Nedum & Nigeria & FT \\ \hline
    P6 & Mahdi & Nigeria & FT \\ \hline
   P7 &  Bogale &  Ethiopia & CT\\ \hline  
    P8 & Jelani & Kenya (Sub-Saharan Africa) & FM  \\ \hline
    P9 & Nathan & Uganda & FM \\ \hline
    P10 & Dinai & Tigray, Ethiopia & FM\\ \hline
    P11 & Pauline & South Africa & FM \\ \hline
    P12 & Samuel & South Africa & FM\\  \hline
     P13 & Fatim & Kenya & FT \\ \hline
    P14 & Kidan & Ethiopia & CT \\ \hline
    P15 & Kevin & Nigeria & FT \\ \hline
    
    \bottomrule
  \end{tabular}
  }
  \caption{Interview participants' pseudonym and market/region. N=15. Gender (8 = male, 7 = female). CT: Current TikTok, FT: Former TikTok, FM: Former Meta}
  \label{participants}
\end{table}

\subsubsection{Interview Procedure}

Each online interview began with moderators sharing information about their background, how they got into content moderation, and their current status. Participants were asked about their experience as content moderators, and the main factors that impact their mental health. These could include working condition, the content, social support, or anything else they considered important. Because interviews included both current and former moderators, \revision{the first author} asked former moderators about their mental health experiences and the impact of the job on their mental well-being, the current support mechanisms. \revision{We} used the moderators' \revision{qualitative} responses in the survey as prompts to expand on the insights they had previously shared with us. Finally, \revision{he} asked moderators for their recommendations \revision{to} improve or support the\revision{ir} mental health. \revision{With the consent of all participants, interviews were recorded and transcribed.} 

\subsubsection{Qualitative Coding}

We conducted an inductive thematic analysis \cite{kiger2020thematic} on the interview data using the software MAXQDA. \revision{The first author and two co-authors were involved in coding interviews. We started our analysis by conducting a line-by-line open coding on five interviews to closely examine the data.} We then discussed our preliminary codes and potential themes and completed open coding for the rest of the interviews.
\revision{We split the interviews among the three researchers and conducted focused coding to identify main themes and organize and revise codes around them. We iteratively discussed with content moderators in our research team through weekly meetings to ensure our interpretation reflect their experiences, and reached into the main themes discussed in this paper. Then, through rounds of discussions with the rest of the research team, we exchanged}
our understanding of the data, refine the identified main themes, and structure the findings. Finally, we conducted member checking with our participants by sharing the summary of the findings and the full draft to ensure the findings reflect their experiences and incorporate their suggestions.

\subsection{Positionality Statement}

Our team's composition and lived experiences have shaped our approach to examining the mental health of African content moderators. Our team includes former content moderators with firsthand experience working in the region under conditions similar to those examined in this study. One of them is also a clinical psychologist who has developed a mental health program to actively aid content moderators. This insider perspective has informed our research design, data collection methods, and interpretation of findings, providing critical context that might otherwise be overlooked in studies conducted by external researchers alone. \revision{Additionally, a majority of the authors are socialized within Africa.} The team brings an interdisciplinary expertise in data work, activism, digital health, and social media research. Our positionality as both researchers and, in some cases, members of the affected community, influences our framing of the study, perspective on the systemic issues we identify, and our recommendations for improving the conditions.


\label{sec:findings}
\section{FINDINGS}

The survey and interview findings reveal multiple factors that moderators associate with their psychological distress and mental health conditions. We present quantitative findings from the survey first (section \ref{survey_mental_health_findings}), followed by qualitative findings that contextualize moderators' experiences with working conditions \revision{(Section \ref{labor_exploitation}), failed corporate wellness programs (Section \ref{failed_wellness}), and long-term impact on moderators' mental health (Section \ref{permanent_trauma})}. 

\begin{figure}
    \centering
    \includegraphics[width=1.0\linewidth]{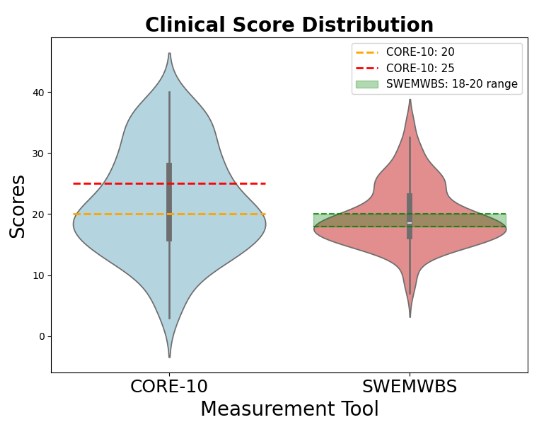}
     \caption{Clinical score distribution of participants across both measurements. 
     CORE-10 >= 25 is severe, while >20 is moderate to severe psychological distress. A SWEMWBS score of 18-20 is indicative of possible mild depression.}
    \label{fig:cinical_score_distribution}
\end{figure}

\subsection{African Content Moderators Mental Health}
\label{survey_mental_health_findings}

We begin with a descriptive overview of the results from the CORE-10 and SWEMWBS measures ($n=134$). The commercial content moderators reported moderate to severe psychological distress with CORE-10 ($\bar x=21.9$, $s$=8.6) and low mental well-being with SWEMWBS ($\bar x =19.8$, $s$=5.1), found in Table~\ref{tab:mental_health_outcomes}. Recall that CORE-10 measures psychological distress, where higher scores indicate more severe distress; on the other hand, SWEMWBS measures positive well-being, so a higher score indicates better well-being. Figure~\ref{fig:cinical_score_distribution} shows the distribution of respondents across the two measures. A CORE-10 score above 20 indicates moderate-to-severe psychological distress\footnote{\url{https://novopsych.com/assessments/outcome-monitoring/clinical-outcomes-in-routine-evaluation-10-core-10/}}.

More than half of moderators ($n=74$, $55\%$) fall into the severe to moderate-to-severe category of psychological distress, and over a third ($n=48$, $35\%$) fall into the severe category of psychological distress. Less than 7\% ($n=9$) fall into the low level or healthy category of distress. 
The analysis of SWEMWEBS supports this analysis -- 52\% (SWEMWBS $<= 18$) of the moderators are in probable clinical depression conditions\footnote{\url{https://warwick.ac.uk/services/innovations/wemwbs/how/}}. In comparison, the general population typically scores at 9.9 on CORE-10~\cite{barkham2013core} and 23.7 on SWEMWBS~\cite{ng_fat_evaluating_2017}. These descriptive results indicate that African content moderators experience negative psychological state and well-being.

Finally, this distress has notable outcomes. About 66\% ($n=88$) of moderators reported that they were exposed to distressing content multiple times a day. 80\% ($n=112$) reported that they were exposed to hate speech, nudity, suicide and self-harm, child abuse, violence incitement, and animal cruelty content. 27.6\% ($n=37$) moderators reported that they are still using drugs and medications to deal with their mental health issues.

\begin{table}[h!]
\centering
\small
\resizebox{\columnwidth}{!}{
\begin{tabular}{@{}llcc@{}}
\toprule
\textbf{Category} & \textbf{Moderator Group} & \textbf{CORE-10} & \textbf{SWEMWBS} \\
                  &                & \textbf{Mean (SD)} & \textbf{Mean (SD)} \\
\midrule
\multirow{2}{*}{\textbf{Location}} 
                  & African moderators        & 21.90 (8.60) & 19.80 (5.10) \\
                  & Global moderators~\cite{spence2025content}   & 10.69 (7.21)***    & 22.66 (5.08)***       \\
\cmidrule(lr){1-4}
\multirow{2}{*}{\textbf{Status}} 
                  & Current moderators       & 17.20 (7.60)** & 21.90 (5.20) \\
                  & Former moderators        & 21.10 (8.40)** & 19.90 (5.00) \\
\bottomrule
\multicolumn{4}{l}{\footnotesize Note: *** $p<0.001$, ** $p<0.05$.}
\end{tabular}
}
\caption{The table presents bivariate comparisons of the CORE-10, Clinical Outcomes in Routine Evaluation 10; SWEMWBS, Short Warwick-Edinburgh Mental Well-being Scale clinical scores for each group. Data from the global comparisons are source from~\citet{spence2025content}.}
\label{tab:mental_health_outcomes}
\end{table}

\subsubsection{Regional Differences}
Next, we compare African moderators' mental health to that of moderators across other regions to understand how moderators' psychological well-being differs across regions, given recent findings from other regions \cite{spence2024content, spence2025content}.
We compare our results to a global survey of the well-being of commercial content moderators by \cite{spence2025content} that replicates and extends their previous work~\cite{spence2024content}. In \citet{spence2025content}, \revision{from a survey of 160 moderators who are from Asia, North America, Europe, and South America, they found an average CORE-10 of 10.69 ($s$ = 7.21) and SWEMWBS of 22.66 ($s$=5.08). This provides an interesting testbed as similar measures are used}. Using a Welch's $t$-test for unequal variance, we find that African content moderators have significantly worse psychological distress ($t=11.97$, $p<0.0001$) and lower well-being ($t=4.79$ $p<0.0001$) than the sample in~\citet{spence2025content} (as shown in Table \ref{tab:mental_health_outcomes}.) Thus, compared to the baseline available, African content moderators' psychological distress and well-being are collectively worse off than global averages of well-being of content moderators.

\subsubsection{\revision{Exploratory Analysis on Factors Impacting Moderators Mental Health}}

\revision{We conducted an exploratory analysis of our survey results using OLS regression to examine both psychological distress (CORE-10) and mental wellbeing (SWEMWBS) (see Methods). The Core-10 model explained 44.3\% of variance (adjusted $R^2 = 0.443$), while the SWEMWBS model explained 32.2\% of variance (adjusted $R^2 = 0.322$). Our exploratory multivariate analysis identified several variables significantly influencing content moderators' mental health outcomes. Working conditions -- such as exposure frequency, duration worked, current employment status as moderators, use of medications/drugs, and, for former moderators, their other employment status -- were found to significantly affect both psychological distress and well-being (See Table \ref{tab:core_10_and_swemwbs_OLS_results}).}

\revision{For psychological distress (CORE-10), the significant predictors were current employment as a moderator ($\beta = -3.749$, $p = 0.011$) -- where current moderators have significantly lower psychological distress (3.7 points lower) than their former counterparts; other current employment ($\beta = -4.099$, $p = 0.004$) -- former moderators who are currently employed in other occupations have significantly lower psychological distress (4.1 points lower) than those who are unemployed; medication (drug) use ($\beta = 6.978$, $p < 0.001$) -- moderators who use medications/drugs show significantly higher psychological distress (7.0 points higher), which could be both a consequence of and response to their psychological distress; and frequency of exposure to harmful content ($\beta = 0.902$, $p = 0.016$) -- highlighting that more frequent exposure is associated with a significant increase in psychological distress.}

\revision{The well-being analysis (SWEMWBS) showed complementary patterns. Worked duration ($\beta = -1.528$, $p = 0.001$) emerged as a significant predictor of lower wellbeing, suggesting that longer tenure as a content moderator is associated with reduced mental wellbeing (1.5 points lower per unit increase in duration). Similarly, medication/drug use ($\beta = -3.592$, $p < 0.001$) was associated with 3.6 points lower well-being scores.}

\revision{To understand the psychological distress across platforms, we look at the average scores of Meta and TikTok moderators. We found moderators at both platforms have higher distress with Meta ($mean = 25.9$, $std = 6.9$) and TikTok ($mean = 21.1$, $std = 8.4$) Core-10 average scores. Moderators at both platforms have low well-being, with Meta moderators' SWEMWBS score ($mean = 18.4$, $std = 4.5$) and TikTok moderators' reporting ($mean = 19.9$, $std = 5.0$). Among Meta moderators, the vast majority ($n = 38$, $77.55$\%) fell into the severe or moderate-to-severe psychological distress categories, while $47$\% of TikTok moderators fall into the same category. $46$\% Meta and $14.7$\% TikTok moderators reported using drugs/medications as a coping mechanism.}

\revisionTable{%
\begin{table}[htbp]
\centering
\revision{
\resizebox{\columnwidth}{!}{
\begin{tabular}{@{}lc@{\hspace{1em}}cc@{\hspace{1em}}c@{}}
\toprule
 & \multicolumn{2}{c}{\textbf{CORE-10}} & \multicolumn{2}{c}{\textbf{SWEMWBS}} \\
\cmidrule(lr){2-3} \cmidrule(lr){4-5}
\textbf{Variable} & \textbf{Coef.} & \textbf{p-val} & \textbf{Coef.} & \textbf{p-val} \\
\cmidrule(r){1-1}\cmidrule(lr){2-3}\cmidrule(l){4-5}
Constant & 14.615*** & 0.000 & 23.563*** & 0.000 \\
Work Duration & 0.911 & 0.198 & -1.528*** & 0.001 \\
Current moderator & -3.749** & 0.011 & 1.773 & 0.068 \\
Other employment & -4.099** & 0.004 & 0.434 & 0.638 \\
Medications & 6.978*** & 0.000 & -3.592*** & 0.000 \\
Months since stopped & 0.058 & 0.294 & -0.013 & 0.721 \\
Exposure frequency & 0.902** & 0.016 & -0.038 & 0.875 \\
Platform: Other & 1.101 & 0.689 & -2.875 & 0.115 \\
Platform: TikTok & -0.645 & 0.704 & -0.113 & 0.920 \\
Gender: Male & -0.568 & 0.628 & 1.446 & 0.063 \\
Gender: Non Binary & -4.354 & 0.507 & -4.999 & 0.250 \\
BPO: Concentrix & -3.715 & 0.199 & 4.754** & 0.014 \\
BPO: Other & 4.075 & 0.235 & -0.521 & 0.818 \\
\cmidrule(r){1-1}\cmidrule(lr){2-3}\cmidrule(l){4-5}
\multicolumn{5}{l}{\textit{Model Statistics}} \\
N & 134 &  & 134 &  \\
R-squared & 0.493 &  & 0.383 &  \\
Adj. R-squared & 0.443 &  & 0.322 &  \\
F-statistic & 9.812 &  & 6.270 &  \\
Prob (F-stat) & 3.80e-13 &  & 1.61e-08 &  \\
\bottomrule
\multicolumn{5}{l}{\footnotesize Note: ***p$<$0.01, **p$<$0.05.} \\
\multicolumn{5}{l}{\footnotesize CORE-10: higher scores = higher distress. SWEMWBS: higher scores = higher wellbeing.} \\
\end{tabular}
}
}
\caption{\revision{OLS Regression Results: Impact of Variables on Psychological Distress (CORE-10) and Mental Wellbeing (SWEMWBS) Scores.}}
\label{tab:core_10_and_swemwbs_OLS_results}
\end{table}
}%

\subsubsection{Long-term Impacts on Moderators' Mental Health} 

Finally, \revision{as former moderators have significantly higher psychological distress while controlling for other confounding variables, we conducted further bivariate analysis to understand differences between these two groups (current vs former) to understand the long-term impact of the job on moderators.} Hence, we compare current moderators ($n=44$) with the former moderators ($n=34$) on TikTok (given incomplete data of current Meta moderators (see section \ref{participant_recruitment})) and their mental health as measured by CORE-10 and SWEMWBS. Table\ref{tab:mental_health_outcomes} reports the average CORE-10 and SWEMWBS scores for current moderators ($CORE-10=17.20$, $SWEMWBS=21.90$) and former moderators ($CORE-10=21.10$, $SWEMWBS=19.90$). Former moderators experience significantly higher psychological distress as reported by CORE-10 compared to current moderators ($t=-2.145$, $p=0.0352$) with moderate effect size ($d=-0.4898$). Differences in SWEMWBS were marginally significant, as measured by Mann-Whitney U ($U=907$, $p=0.1097$). About 32\% ($n=14$) of current moderators reported a CORE-10 score in moderate psychological distress level, and another 32\% ($n=14$) in severe and moderate-to-severe categories. In comparison, nearly half (47\%, $n=16$) of former moderators were in severe or moderate-to-severe psychological distress. This shows a significant difference in psychological distress between these two groups of moderators, and former moderators fare worse than current moderators. We explore why this is the case in our qualitative findings below.

\subsection{\revision{Interviews:} Enmeshed in Precarious Working Conditions} \label{labor_exploitation} 

\revision{Next, we transition to the results of our interview study. B}eginning from the recruitment stage to post-employment, \revision{the working conditions} contributes to severe mental health conditions identified in our survey (Section \ref{survey_mental_health_findings}). Moderators often enter this work without being informed of the risks. Many \revision{must relocate for work }
but later face \revision{legal issues due to} expired work permit. Their meager salary 
prevent\revision{s} workers from \revision{seeking} necessary mental health support outside BPOs' offerings. \revision{Besides the} exposure to disturbing content, the toxic working environment adds to their psychological stress. However, due to Non-Disclosure Agreements (NDAs), social stigma, and safety concerns, moderators remain silent about their experiences. 
We discuss these systemic factors as follows. \revision{In this section, we refer to participants by their pseudonyms. Quotes have been lightly edited for readability.}

\subsubsection{Lack of Transparency in Recruitment and Onboarding}

Moderators recalled that the job description did not \revision{honestly disclose the disturbing} tasks involved \revision{-- j}ob descriptions sometimes misleadingly included terms like ``customer service'' and ``translation.'' Many moderators only learned about the job responsibilities during \revision{onboarding}. \revision{As Nedum-FT pointed out: \textit{``we were deceived and tricked...initially from the recruitment process until the time of signing the contract, we don't know [the actual job tasks] until the first day we begin training.''}} This lack of transparency also extends to onboarding, as the content provided for training was significantly less disturbing than the content moderators encounter while on the job. \revision{Zeyneb-FT mentioned that during training, \textit{``they were using simple videos that were not violent.''}}.

Opaque and deceptive recruitment can have serious consequences for workers. \revision{M}any moderators \revision{moved} to different countries for this job, and found it \revision{prohibitively} expensive to \revision{move} again after realizing the actual tasks of the work. Moreover, the recruitment process fail\revision{ed} to assess applicants' current mental well-being. As a result, moderators with prior trauma \revision{(}who are not suitable to work with disturbing content\revision{) were} inadequately informed of the real job requirements. 
Mahdi-FT highlighted the importance of understanding moderation work before \revision{beginning}: 
\begin{quote}
    ``If there is clear explanation of...responsibility [of this work], it's fine...[But] some people may have some mental issues...and then if they join the job [without knowing], it's gonna be a different story.''
\end{quote}

\subsubsection{Expired Work Permits}
\revision{Most} moderators are hired from different countries and move to the workstations (i.e., Kenya, Ghana, Tunisia) where the BPOs operate. They began \revision{working} with short-term work permits that were significantly shorter than the duration of their contracts. These permits were supposed to be transitioned into longer-term work permits after BPOs completed background checks, training, and other assessments. However, longer-term work permits remain pending and in process, leaving moderators with expired work permits. \revision{Badru-CT described his frustrating experience to get a longer permit, suggesting that the BPO intentionally did not update their permits: \textit{``I've been at the Kenyan immigration office on several occasions. And each time I go...the documents required to process the work permit were not being uploaded accurately and adequately. And when you don't do that, how do you expect the government to provide the work permit?''.} }
Invalid work permits put moderators under constant anxiety and fear. Badru-CT described how this affects him, especially when facing law enforcement: 

\begin{quote}
    ``When I want to go to shop...if [a police officer] stopped me, he will ask me for money unless I give him my ID card...It is a trouble for us every time when [I] encountered these things, somehow it's affecting me. When I see a police officer, my heart will fail.'' 
\end{quote}

These invalid work permits also restrict moderators from visiting famil\revision{y} and friends in their home countries. Although moderators are entitled to an annual trip back to their country, financially covered by the BPOs,  invalid work permits render this option basically impossible. 
Shuman-CT told us that a Nigerian colleague committed suicide due to being unable to return to her home country: 
\begin{quote}
    ``For a while she has been trying to go back home, but then it wasn't possible because of certain restriction issues with her [work] permit. Then she didn't come to work for a while. Now we know that she died.''
\end{quote}

\subsubsection{Constraints on \revision{S}peaking \revision{O}ut about \revision{W}ork} \label{constraints}
\revision{O}rganizational and societal factors also prevent moderators from speaking about their experiences and harm their well-being. Non-Disclosure Agreements (NDAs) prohibit them from sharing their work experiences, including the identity of their clients (i.e., social media platforms), the location of the office, \revision{and} the types of content they work on. \revision{Nathan-FM pointed out the consequence of signing NDAs: \textit{``as soon as you're hired...you sign an NDA to not talk about anything to do with content moderation.''} Shuman-CT further compared how NDAs are silencing them metaphorically to physical violence, \textit{``it's as if you are beating us, and then we can't even seek help [because of the NDA]. We can't really express ourselves out there''}} 

Stigma associated with graphic, violent, and especially adult content prevents them from \revision{also} discussing their experiences. Zeyneb-FT explained: \textit{``you cannot talk to someone about [the work], because they are going to judge you, and they wouldn't care if it's your job, they will just say you watch such content.''} The personal safety concerns and fears of being targeted for their jobs, especially in regions with high conflict rates, are a great concern, adding to the mental strain. Nathan-FM attributed these concerns to a lack of public awareness \revision{about} moderation, which \revision{partly} results from strict NDAs: 
\begin{quote}
    
    \revision{``I pretty much think it's because people didn't know what we do or why we do what we do (...) the public doesn't know [this job]. And maybe they should make [NDA] a bit less protected like that.''} 
\end{quote}

\subsubsection{Poor Compensation} \label{low_pay}

Moderators described \revision{that the} financial instability \revision{made it difficult to meet basic needs and support their families, significantly}
affecting their well-being. 
Workers who were initially promised salaries sufficient to cover their living expenses faced the opposite reality of inadequate wages amid high living costs. Badru-CT stated, \textit{``Every time I have rent, a bill to pay...it is like they are taking my life for nothing.''} Moderators also reported that low pay prevented them from accessing mental health services outside corporate programs. Zeyneb-FT explained, \textit{``Most people don't have the money, they don't go to outside counsellors.''} Moderators also described unexpected contract terminations. \revision{Shuman-CT noted a recent case of contract termination where \textit{``you have people, two days into the end of their contract, you are telling them that it will not be renewed.''}}.

\subsubsection{Toxic Work Environment}
\label{toxic_work_enviornment}
The toxic work environment significantly adds to the mental strain. Moderators face a heavy workload and intense surveillance under algorithmic management and their managers. Utilization rate, or how often moderators are using the system,  is one of the metrics that workers need to keep up, \revision{and} 
if \revision{it} goes below a certain level, Bogale-CT mentioned that \textit{``the management will send you word, name letters.''} Bogale gave an example of what this looks like: \textit{``[If] I'm told to [review] like 20 reported videos per hour. So I completed these videos...[In] the rest of the hour, there is no video going to come, [but] I have to open an empty queue, and then have to refresh the system [to keep utilization rate high].''} 

Furthermore, moderators have minimal breaks, and even so, the breaks are not well respected by BPOs. Tanisha-CT noted the difference of how breaks are conceived by clients (Meta/TikTok) and implemented by BPOs: 
\begin{quote}
    ``When you're going through training on policies, you see the clients [Meta/TikTok] always emphasizing wellness.... but when it comes to doing the job [with the BPO], these things are practically not there. It's just working, working, working. Nobody cares what's happening to these people.''
\end{quote}

Even during breaks, moderators are prohibited from using social media platforms for entertainment\revision{, which denies them opportunities to relax after dealing with extremely graphic content and tight work situations.} 
Nedum-FT described this experience:

\begin{quote}
    ``When you are working you are tired, or you see some content that is bad...when you're on break, you can maybe watch some YouTube, or maybe, if you are a football lover, you can watch some highlights or music, and you'll be happy, right? But they block YouTube, block the links you can't even access anything on your system other than the content moderation platform.''
\end{quote}

Even when client/social media representatives visit the company, moderators have very limited opportunities to voice their concerns. BPOs select specific employees who are silent to engage in discussions with the clients, so the extreme working conditions are not brought up. 
Zeyneb-FT described a site visit by a client: 
\begin{quote}
    ``when the client is visiting, even if I [go] into the toilet, and maybe the project manager will follow [the representative] to the toilet so that he can't speak to anyone...They wouldn't leave the client by himself.''
\end{quote}

Some moderators hope that clients can intervene after being fully informed about the working conditions. As Tanisha-CT puts: \textit{``if the [client] company truly wants to do the right and wants to maintain its image...the company should be able to do that, finding a way to carry that into an investigation [into BPOs' practices].''} However, others are less optimistic, as communicating directly with client representatives also does not guarantee that the concerns will be meaningfully addressed. Badru-CT mentioned an instance of directly reporting concerns to client representatives: \textit{``Nothing will happen. The next day, nothing. Everything is the same.''} He further mentioned a meeting where people from the headquarters were present: \textit{``I heard the company head [said] in one meeting, if you don't like the salary we are paying you, you can find another job, sir.''} 

\subsection{Failed Corporate Wellness Programs} \label{failed_wellness}

\revision{African content moderators have severe mental strain, yet their working conditions do not support them to cope with these issues, but instead intensifies their stress and struggles. Corporate wellness programs are a primary, if not the only, resource for moderators to address their mental health challenges. Companies offer corporate wellness programs to support moderators, including individual and group wellness and therapy sessions, and group activities. However, our study revealed that all participants found them unhelpful or ineffective. We outline the reasons workers perceive as causing the ineffectiveness or even failure of corporate wellness programs.}

\subsubsection{Lack of Confidentiality in Sessions} \label{confidentiality}
All moderators express a lack of confidentiality as a main reason for them to abandon the wellness counseling or therapy sessions. Privacy and confidentiality are at the core of therapy sessions, where sensitive individual information is kept private between the parties involved. They expressed instances where issues discussed in their 1-1 sessions were being discussed outside and making their way to the management and human resources. This inhibits moderators from freely expressing when seeking mental support. 
As Shuman-CT put it:

\begin{quote}
``we are not even too confident with the counselors we have...Whatever we share, either personally or as it relates to the job, will get to the ears of the managers, the ears of fellow colleagues.'' 
\end{quote}

\revision{Tanisha-CT mentioned that \textit{``somebody once told me that something like that once leaked out. So [from] that point on, he started avoiding speaking to therapist about confidential issues.''}} The sensitive information can put them at risk of being blacklisted by their supervisors and management. Zeyneb-FT describes, \textit{``[the therapists] release such information to the management, and then you get victimized...that has affected the way we relate to one another. We have people that are really tied down emotionally. You have people that [have] become less expressive.''} 

\subsubsection{Lack of Expertise in Handling Traumatic Experiences} 
Our participants also questioned the qualifications and experience of the available counselors provided by BPOs. They mentioned \revision{the inexperience of therapists, such as having superficial and} abstract-level discussions\revision{,} 
and \revision{raising} trauma-triggering questions. 
\revision{For example, Nathan-FM who has trypophobia mentioned that `\textit{`the counselors didn't know what trypophobia is. It was like a new thing to them. And for me, I wouldn't show them what it was, because I felt it was bad for me, imagine showing it to a person who has the same problem.''} Jelani mentioned that \textit{``[you] explain what you just saw to someone, and they'll just tell you, everything will be okay, nothing else.''}}
The moderators expressed feeling no change in their state of mind, and sometimes even re-live their traumatic experience in these sessions. Kidan-CT told us:

\begin{quote}
``The wellness [therapists], they'll be like, how are you doing today? So, what are you planning for your future? I'm like, how is my future?... Where does my future come in? I'm actually in pain right now.'' 
\end{quote}
Kidan-CT recalled a particularly painful experience with a corporate wellness therapist:

\begin{quote}

    ``The therapist [was] trying to explain the image I had seen. [Kidan recounts a graphic description of the content]. The therapist is asking me, What content make you feel, or what was the worst part like? He'll start making me imagine the content and start feeling like, why did this person do this?''
\end{quote}

Moderation is a job where moderators are faced with unfiltered content on these platforms. Recall that NDAs prevent people from discussing this content with others outside the BPO, where they can not seek support from their families and friends, making these in-house corporate wellness programs crucial. Thus, the moderators felt a disconnect between the reality they express during these sessions and the abstract responses they receive, which do not address their state of mind.

\subsubsection{Ineffective Wellness Activities}: In addition to individual counseling services, the BPOs organize group wellness activities. These activities, aimed at team bonding and wellness, include games between groups of moderators and group counseling sessions. However, 
the tense and toxic work environment (see Section \ref{toxic_work_enviornment}) and unhealthy relationship with the management lead moderators to not embrace these activities. Moderators also felt the disconnect and pressure to attend these sessions. Moreover, the lack of creativity in these sessions contributed less value in positively changing the mental status of moderators. Shuman-CT expresses the effectiveness of these sessions:

\begin{quote}
    ``Initially, when we came we had like two [group sessions]. I went to one, and it was actually fun, because we just came in. The environment wasn't these tense. So people actually came to it. But for now, people don't really care.'' 
\end{quote}

Dinai-FM also found this frustrating: {\it ``They get a game at that time. We don't want to play any game, you know -- we are stressed.''} \revision{Shuman-CT mentioned that {\it ``sometimes those team building are organized during weekends (...) and for me, I basically want to use the weekend to rest.}'' He described how these activities increase tension: 
\begin{quote}
``the people that turn up for such [activities] is very limited. Now...they're making it compulsory for people to go, whatever is being budgeted for it, whatever expenses will be deducted from your salary. But now people don't care. They say, if you want to deduct, [then] deduct, but you cannot force me go there.''
\end{quote}
}

\subsubsection{Management Fails to Implement Counselor Recommendations} The companies also failed to implement recommendations from counselors and therapists. Counselors and therapists carry a responsibility beyond assisting moderators: they must identify necessary changes in BPO practices that affect moderators' mental health conditions. Some therapists and counselors who genuinely cared for and understood moderators' dire mental health situations reported their recommendations to higher management to improve working conditions, but management did not implement these changes. Nedum-FT describes:

\begin{quote}
    ``(my therapist says), I passed your message across to the head of the Wellness department, the therapist lead. I'm waiting for a response, because she also needs to pass the message up. And you see that, right? So I'm saying they also have their limit right now, the guys who could actually make this thing happen, who could make your work permits come to life, are the guys at the very top. Now you find that, even the therapists that are going to speak about it, they're also going to try and be careful so they don't find themselves in the black bad book."
\end{quote}

Other factors, such as \revision{few therapists,} limited access to therapy sessions, and their restricted duration compound these issues. \revision{During our interview, Tanisha-CT quickly calculated the availability of therapists in relation to the number of moderators: 
\begin{quote}
    ``[You have] over 200 employees on shift at the same time. There is a maximum of two therapists, [and] each therapist can spend either 20 or 30 minutes with employees...Now a therapist works for eight hours [with] one hour break... so one therapist can see maximum of 21 employees a day. In total, that's 42 for over 200 employees. So that's just around 20\%.''
\end{quote}
}
\revision{Nedum-FT mentioned that \textit{``most people don't even want to go there [wellness sessions], because it's no use. So the team manager will...force you [to go], just because it's a process.''}}  Existing corporate wellness programs, therapy sessions, and group activities fail to deliver their intended services. While these corporate-provided programs may appear comprehensive on paper and highly advocated for by BPOs and social media platforms when confronted with moderators' mental health, they currently serve only as superficial gestures rather than addressing moderators' urgent mental health needs.

\subsection{Permanent Trauma and Mental Health Damage Post Moderation}\label{permanent_trauma}

Finally, the trauma of content moderation persists long-term, continuing for months and years after moderators leave their positions. Consequences include psychological and behavioral changes that damage relationships with family and friends. Moderators report that the job's impact becomes even more severe after leaving moderation work, as they must now face the real world without the coping mechanisms developed for that artificial environment. Our clinical survey results \revision{confirms this, showing }
that former moderators experience significantly higher psychological distress and lower mental well-being than current moderators (Section \ref{survey_mental_health_findings}). This finding highlights that beyond the systemic failures in supporting current moderators, companies must provide extensive attention and support for workers after their moderation careers end.

\subsubsection{Psychological Distress Consequences on Social Life} Former moderators often express impacts of their work on personal, family, and social factors. They discuss long-term effects on their mental health and drastic changes in their mood and behavior. As workers were not able to process their emotions in a timely manner but had to suppress them, they developed psychological disorders such as social anxiety, and headaches, stress, isolation, and frustration. These disorders later manifest into behavioral changes such as avoidance, self-isolation, and withdrawal from friends and family, which can worsen their mental health conditions.

Daily exposure to harmful content also transforms how moderators view society. Moderators express feeling helpless within the society they inhabit, imagining that individuals around them might commit the actions they witnessed in videos. 
\revision{As Nedum-FT described: \textit{``you don't want to be close to your family and friends you see...you will be picturing them as if they were something you see on the content [you moderated].''}} 
Moderators experience this shift in perspective during daily activities, leading to them becoming extremely protective of their children, struggling with intimacy due to constant exposure to sexual content, and feeling anxious when speaking to strangers on public transportation. Jelani's-FM experience summarizes these challenges faced at a personal, family, and societal level.

\begin{quote}
    ``I'm no longer interested in life. I'm a father to a young girl. She's turning three years in the next two to three months, and I found myself quite overprotective of her. Just because of what I came across (...). I'm afraid every time I'm on the road, I'm just thinking of how I'll get from point A to point B, without being involved in a fatal accident. I found myself afraid of blood (...). Occasionally, I might have flashbacks of what we used to do, something that I find is also affecting my family, because now in such cases. I have to distance myself from my kid and my partner, just so I can have the thoughts to myself only without involving them. Because I feel that if I start talking about what I am experiencing, it also might affect them in a bad way.''
\end{quote}

\revision{Similarly, Pauline-FM expressed her anxiety and fear after moderating extensive content related to gender-based violence: 
\begin{quote}
``There was a time where there was a lot of gender based violence going on...it's a crisis in South Africa. I never used to mind going for a walk in the evening, but as I came back [after work], I don't even use Uber anymore, because I can't trust anyone. I had to remove some of the graphic content of women being raped, a woman being killed...I am a very like outdoor person, but I don't do that anymore.''
\end{quote}
}

These permanent traumas extend beyond psychological damage to include physical harm. Back pain, \revision{``weak'' and ``blurry'' vision,} and hearing damage rank among the most common physical issues. \revision{For example, Jelani-FM mentioned that \textit{``due to the long periods that we were sitting down, I experience back pains once in a while...And it's not just me. You'll find a couple of guys also complaining about back pain.''}}Extended work hours, unhealthy equipment, and surveillance-focused environmental setups inflict these physical repercussions on moderators. Insomnia and constant nightmares also create physical health problems through chronic exhaustion.

\subsubsection{Normalization of Extreme Behaviors} Moderators are frequently exposed to extremely disturbing content that become embedded in their subconscious mind. \revision{Samuel-FM puts: \textit{``it makes you a different person. (...) Seeing all these things becomes normal to you, and it's things that should never be normal to anybody.'' }} Moderators described the distressing inability to control thoughts of extreme behaviors that come to mind repeatedly. Kidan-CT states:

\begin{quote}
    ``It actually plays in your mind. I will be sitting down and I'm looking at something and I'm imagining, oh, I can do this. (...) 
    The more you continue doing these things, the more you have no support, the more you have no one to speak to, that voice starts to die, and it reaches a point where someone can kill himself, or maybe starts doing like suicide\revision{.}'' 
\end{quote}

\revision{Participants mentioned dissociation during work as coping mechanisms. For instance, Mahdi-FT indicated disconnecting from oneself at work: \textit{``most times you are working, you don't really have your own self again, you're in another self.''} Another example is that Nathan-FM indicated feeling the content was unreal: \textit{``you're doing your job...if you see someone crashed on the road, it's just funny, it's not real. It's like a coping mechanism.''}}
Some moderators \revision{indicated relying on drugs to cope with extreme content during the work} 
and a systematically failed support mechanism (section \ref{failed_wellness}). 
\revision{For example, Fatim-FT mentioned: \textit{``I had several friends...they use drugs to try to be off from getting to see these things [content] over and over again.'' Similarly, Pauline-FM said: ``a lot of us resorted to using alcohol and drugs to just forget and survive.''}} 


\subsubsection{Effects on Career Aspects} 

Moderators face long-term employment challenges after their content moderation careers end. Commercial content moderation exists only from social media platforms such as Facebook and TikTok, and are provided through select BPOs in a few countries like Kenya, Nigeria, Ghana, and Tunisia. Outside of these specific circumstances, participants felt that their job experience as a moderator becomes largely irrelevant for other job prospects across Africa. After spending years in the industry, moderators must re-enter the job market while managing additional mental health instability. This career gap, combined with economic challenges, makes finding new employment opportunities extremely difficult. Nathan-FM describes:

\begin{quote}
    I have to explain to \revision{[future employers]} what that is. And most of the times that never it is not something that is relevant in the job market (...) Which job are you really doing? You know you're not doing any other thing.'' 
\end{quote}

\revision{He further pointed out that the labor organizing activities affected their future job applications even within the field of content moderation: 
\textit{``whenever I applied, I wouldn't get a job because they know I was blacklisted...[if] I want to stay in moderation, they wouldn't hire someone who sued their previous employer.''}} 

After moderators leave their work countries (such as Kenya or Ghana) and return home, companies provide only an international hotline to contact therapists in Kenya during times of distress. These hotlines was ineffective due to the high costs of international calling. \revision{Pauline-FM explained: \textit{``They just gave us a number to call, and they didn't even consider the fact that I am in South Africa. I don't have air time to call that Kenyan number, and I'm using a South African number. How will I be able to get in touch with those people [therapists]?'' }}This represents another component of the systemic failure within corporate wellness programs that appear reasonable on paper but are impractical in reality (section \ref{failed_wellness}).


\subsection{Towards Better Mental Health Experience for Content Moderators}
Moderators envisioned different ways to improve their work experiences and mental well-being. These proposals include technical interventions, better mental health support, and improved labor conditions, along with greater social recognition. 
\subsubsection{Designing and Implementing Technical Interventions}
Moderators discussed different technical interventions to improve working experiences. Several moderators reported that an existing technical feature, which marks partial images gray to prevent exposure to disturbing scenes, is ineffective. Zeyneb-FT explained the problem when using this feature: \textit{``you're going to miss a violation, because the image should be very clear for you to type any of the violation there. Then, you will not get the accuracy correct.''} 

Instead, moderators envisioned developing and implementing automated moderation for extremely challenging content, which can reduce human exposure to it. According to Jelani, these tools should divide \textit{``what should be handled by human beings and by machines, because we have instances of tickets that should not be handled by human beings at all. They need to have systems in place [to] let [the computer] action [such content] automatically.''}

Moreover, moderators \revision{recommended} using AI to filter traumatizing content by different categories, such as identifying suicidal content, violent content, accidents, and sexual content. This is what is called {\it advanced task routing}, with warning to the moderators about what was in the queue. Moderators proposed that this would help them set their expectations so they are prepared mentally for future content instead of not knowing what is coming. Despite the potential of AI to assist here, moderators were skeptical that AI could do all the work of the moderators. Mahdi-FT reflected:
\begin{quote}
    ``I don't think AI will be 100\% relied on...even though they are not 100\% relied on, those dangerous queue should be segregated so that... few [moderators] will just go to check on it, and they have much break (...) have special treatment because of the kind of queue they are working on.''
\end{quote}

\subsubsection{Improving Mental Health Support}
Moderators identified opportunities to improve wellness program. 
Most moderators agreed about the importance of skilled therapists and mental health professionals -- in particular, those who are \revision{\textit{``trained with trauma'' (Jelani-FM)}} -- and ensuring sufficient counseling hours. Moderators also suggested holding regular check-in therapy sessions rather than offering them on demand\revision{:\textit{``the therapy should not just check up on you work. Even during your off days, they may call you once or twice in a week. How are you doing? You have anything going on with you? Can you share anything with us?'' (Mahdi-FT)}} 

Due to concerns about trust and confidentiality (see Section \ref{confidentiality}), our participants specifically recommended having independent therapists from external companies. Unlike those who work within BPOs, external therapists do not fear managerial retaliation in BPOs and can ensure the confidentiality of the conversations. This can create a safe environment for moderators to open up during therapy sessions and discuss work-related challenges, without fear of retaliation caused by leaked therapy conversations. Shuman-CT shared his thoughts of ``outside'' in different aspects:  
\begin{quote}
    ``that issue of the confidentiality of the entire process needs to be [there]....You have sessions outside...you have counselors outside...I can just choose to take them to outdoor events. It builds confidence. It sticks away that fear, that rigid nature, it makes people free to express themselves.''
\end{quote}

Many also emphasized the need for honest and transparent recruitment, viewing mental strength as an important criterion for this work. Mahdi-FT argued that \textit{``during that recruitment, [applicants] should be very clear that this is the content that [they] will be seeing [in the work]. And then in the mental health screening, they [can] try to identify some people who do not have prior trauma that can be triggered''.}
Mahdi-FT further suggested establishing resilience programs to complement therapies so that therapists know more about moderators and the challenges, and thereby \textit{``how to support [moderators] to be resilient to their mental trauma.''} However, Badru-CT also pointed out the limitations of therapy, such as addressing concerns at workplace and introducing changes: 
\begin{quote}
    ``we are looking for a change...when I tell [the therapist] my problem, who will deal with my problem?...I go with a problem, I want to be answered for that problem...not just hearing, but trying to address that problem.''
\end{quote}

Moreover, moderators often leave the job with mental strains. As Shuman-CT puts it, \textit{``you go with your craziness, you go with your madness, you go with your mental disorders and everything.''} \revision{Fatim-FT emphasized the importance of continued therapies after leaving the job because \textit{``[we] don't have means to get a personal psychologist (...) To have that psychologist to talk is very important because once we are out of the workplace, we need to live a normal life once again.''}} Thus, moderators wanted access to continued mental health support for some time after leaving the job for six months to 12 months. This access could be provided by the BPOs or through an improved insurance policy to allow access to mental health support over an extended period. 

\subsubsection{Creating Better Working Conditions and Fostering Social Recognition} 

Moderators called for improvements in working conditions. Fair compensation surfaced as a significant concern among moderators. Receiving fair pay would help them gain the financial freedom to seek therapies beyond what BPOs offer. 
Jelani-FM also argued that fair compensation can make her \textit{``feel a sense of dignity within you [knowing] this company is treating me as a human being,'' especially considering the pay disparity that African moderators ``are paid significantly less from other moderators [in other countries], and yet are doing the same thing.''} To emphasize the importance of adequate and fair compensation, Nedum-FT points \textit{``Low compensation relative to the emotional and psychological demands of the job should be treated not just as a background condition, but as a central component of the political and economic structures shaping moderators’ well-being."}  
Additionally, moderators emphasized the need for permission to work remotely, and ethical workload planning, such as having ``regular breaks'' and ``reduc[ing] the prolonged exposure [to] violent graphic content.'' \revision{Fatim-FT highlighted the importance of a supportive working environment: \textit{``[moderators] are already having pressure from their content (...) let me deal with the content moderation, but at least the working environment should give me peace.'' }}

Furthermore, moderators often face sudden job termination (Section \ref{low_pay}), which, together with \revision{little saving and} mental strain they have during the job, leaves them struggling in isolation. \revision{As Nathan-FM described, \textit{``you are living in another country. You're paying bills, you're not saving anything,'' and losing this job means ``losing everything all at once''.}}
Therefore, moderators requested structured support to help them transition smoothly when they leave their jobs. Shuman-CT noted, \textit{``[BPOs] should organize certain things to help me transition easily into the world out there.''} 
This transition can include early notice for contract termination, such as two to three months, to give moderators, \revision{especially those relocated in a different country to do the work,} sufficient time to \revision{\textit{``get something back home [and] prepare myself.''} (Shuman-CT).} 
The transition also has professional aspects, as moderators face career gaps that undermine their chances for future employment. For example, 
Shuman-CT envisioned professionalizing content moderation:
\begin{quote}
    ``[content moderation] should be like a profession. It should be something you can grow... because I won't use it in [my] CV, because employers out there, the moment you mention content moderating, it seems like, you're already gone mentally...So if it's something that can be done at the global level, to make it like a career, something professional, it's good.'' 
\end{quote}

Moderators further advocated for relaxing NDAs and increasing social awareness and recognition for this job. 
As Samuel-FM explained, \textit{``everything a normal social media user would not [be] exposed to [is] because of the work content moderators do. For example, things like pornography, hate speech, violence incitement.''} 
This recognition can also help moderators overcome the social stigma and safety concerns that keep them silent (Section \ref{constraints}) and gain more support from social networks.  
Nathan-FM pointed to the connection between public awareness of this job and mental support:
\begin{quote}
    ``
    if people could know how much I protect them as a content moderator, I don't really think I would struggle to open up to my mom. (...) I think mental health support doesn't entirely come from the professionals. It also comes from the support system that you have that is your close family, your friends.'' 
\end{quote}


\section{DISCUSSION}

Our findings highlight the severe mental health conditions of content moderators in Africa and reveal the systemic \revision{factors, beyond frequent exposure to extremely graphic content, that drive this crisis.} 
From recruitment and onboarding---where moderators are attracted \revision{through} opaque and misleading job descriptions---to the maintenance of extremely toxic work environments, BPOs perpetuate harmful conditions. Failed corporate wellness programs further exacerbate moderators' deteriorating mental health, leading to long-term trauma that affects their lives well beyond their employment period. 
\revision{These factors are found across multiple BPOs and platforms and go beyond specific organizational rules. Importantly, they are interconnected and should be situated in broader socio-economic contexts such as unemployment issues, governments' priorities to combat unemployment and boost economic growth at the expense of inadequate protection of workers' well-being and labor conditions, and the lack of public awareness of moderators' work and contributions to the online environment.}

By emphasizing how these factors systematically hinder workers from getting the vital mental health support they need, we argue that addressing moderators' severe mental health challenges requires holistic approaches that go beyond technical fixes or changes in organizational rules and involve diverse stakeholders. 
In the following sections, we discuss structural and policy failures, examine the tensions researchers should consider when engaging with moderators, the need to go beyond design implications, and present actionable recommendations for the directly involved stakeholders.
\subsection{Unique Content Moderation Infrastructure Failures in the African Context}

Comparing our findings with two studies examining moderators primarily across Europe and Asia~\cite{spence2024content, spence2025content}, we found African moderators exhibit higher psychological distress and poorer mental well\revision{-}being. This raises a critical question: why do African moderators experience worse mental health outcomes? What protective policies exist in other regions that are absent in Africa?

Our interviews revealed platforms and BPOs force moderators to operate under extreme conditions (Section \ref{labor_exploitation}) and ineffective corporate wellness programs (Section \ref{failed_wellness}), further exacerbating their mental health issues. In addition, poor government policies in Africa contribute to these problems. Several examples illustrate this disparity. EU countries mandate minimum wage protections for workers. US moderators can skip graphic content and, after legal battles, receive compensation for platform-related harm; in 2020, a court awarded \$52 million to thousands of Meta moderators \cite{allyn_settlement_2020, newton_facebook_2020} and increased US moderator pay~\cite{klonick_for_2019, newton_facebook_2019}. Moderators in the Philippines and Estonia are properly informed about job implications and working conditions before employment \cite{savio2023social, bharucha2023content}. These policies demonstrate other regions have established rules and oversight mechanisms to protect moderation workers. While conditions in these regions remain imperfect and likely below average compared to other occupations, they provide baseline worker protection.

In contrast, African governments actively court BPO companies, viewing the sector as a primary solution to high unemployment and economic growth driver \cite{hoppner_africas_2025}. Kenya's Vision 2030 positions the BPO sector as a key pillar for economic growth \cite{kenya2030vision_economic_nodate}. In spring 2024, Kenyan President William Ruto announced infrastructure expansion plans to attract more BPOs, aiming to create one million jobs over five years \cite{apofeed_kenya_2024}. The mobility of BPO operations creates pressure, as countries fear losing these opportunities despite poor working conditions. Recently, Meta relocated its content moderation operations and BPO contractor from Kenya to Ghana, where moderators face worse conditions \cite{wilmont_metas_2025}. This economic context, where governments prioritize BPO attraction while overlooking negative impacts on citizens, combined with operational flexibility of platforms and BPOs, contributes to substandard working conditions and poor accountability. Consequently, this hostile work environment and inhumane treatment significantly deteriorate moderators' mental health compared to counterparts in other regions.

\subsection{Considerations In Designing and Engaging with Content Moderators}
\label{tensions_in_design_and_engagment_discussion}


Next, we consider several tensions and potential solutions when engaging and designing with commercial content moderators on research and in solutions. Researchers at CHI and beyond have designed technological interventions (i.e., systems that hide or blur graphic content at different levels \cite{das_fast_2020, dang2018but, karunakaran2019testing}), proposing automated detection methods, or even psychological interventions to alleviate this graphic content problem. 
However, moderators in our study highlighted the limitations of current technical interventions in the workplace and stressed the importance of \revision{integrating }interventions into existing workflows and address\revision{ing} \revision{workers'} challenges, which shows the value of designing in close collaboration with the affected communities  \cite{cooper2022systematic, chen2022trauma}. Moderators generally (and specifically in Africa) are marginalized -- if HCI wanted to address these issues through research, what are the tensions that will emerge?

Reflecting on our study, there are several considerations when engaging with content moderat\revision{ors} in these circumstances. First, as argued by \citet{miceli2025methodological}, is the balance between recognizing workers' {\it epistemic authority} and being mindful of {\it epistemic burden}. Researchers should avoid simply extracting knowledge about moderators' experiences. Instead, they should view moderators as collaborators and acknowledge them as experts in their own practices and experiences \cite{dair_dair_2022}. This can be achieved, for example, by treating moderators as co-researchers and inviting them to share their experiences and perspectives in meaningful ways during events. 
At the same time, researchers should be aware of the epistemic burden, which \citet{pierre2021getting} described as burdening oppressed groups with educating privileged groups. 
\revision{M}ental health stigma and the restrictions imposed by NDAs can prevent workers from openly discussing their challenges and experiences. In particular, NDAs create  \revision{create legal risks for moderators to speak up in research and public events, }
even after leaving their positions. 
This again highlights the urgent need to revise NDAs to enable workers to share their experiences without fear of repercussion.

\revision{S}everal existing methods within the CHI community can assist with navigating the challenges of epistemic authority vs burden. These methods center on co-designing \cite{myers1998brief} with communities, such as participatory design \cite{muller1993participatory} and value-sensitive design \cite{friedman1996value}. Other frameworks, such as trauma-informed computing approaches, can better inform how we engage with moderators when designing better technological interventions \cite{chen2022trauma}. Beyond that, how should CHI and the research community engage with commercial moderators? How can we create research practices that center moderators' agency and epistemic authority? What communities of practice can we engage with to prioritize their experiences while protecting their privacy? How can we effectively advocate for content moderators to challenge the structural and systemic problems we have identified?  Researchers might reflect on these questions when they are developing projects involving content moderators.

\subsection{Beyond Design Implications - Collective Movements for Moderator Mental Health Reform}

\revision{Our findings highlight an important tension: while our participants suggested better-designed technological interventions could improve their working experience (e.g., advanced task routing), technological solutions alone may not address the systemic problems of commercial content moderation in Africa. These problems are structural: deceptive labor practices and hiring, poorly trained therapists with no confidentiality, and no support after they finish employment for their ongoing mental health struggles.}

\revision{While Dourish critiqued HCI's need for design implications nearly two decades ago, the risks of envisioning design implications is salient for commercial content moderation in Africa~\cite{dourish_implications_2006}. As~\citet{irani_postcolonial_2010} notes, ``development regimes have historically been aligned with the interests of politically powerful commercial and capital market actors.'' In our case, prioritizing design implications risks serving ``capital market actors'' (tech companies and the BPOs), as it is easier to improve the AI technical system than addressing structural factors. Furthermore, design implications may also erase our participants' perspectives, where this labor is already affected by the invisibility of the work~\cite{gray2019ghost}. Thus, we join with decolonial scholarship about the risk that design implications beyond our participants' own requests could risk exploiting their struggle for performing HCI articulation and research~\cite{irani_postcolonial_2010, pendse_treatment_2022,shahid_decolonizing_2023}.}


\revision{Towards this end, 
we propose social and policy implications that center our participants' lived experiences. We focus on the stakeholders with the power to enact structural change --- BPOs, platforms, governments, and society --- and identify actionable steps that may move the needle in working conditions for moderators.}

\begin{enumerate}
   
    \item \revision{As direct employers, \textbf{BPOs} are central to improving moderators’ working conditions; they can implement regular (i.e., quarterly) mental-health assessments by external clinicians and adjust workloads accordingly, including reducing exposure for workers who show distress. Recruitment should be transparent by showing applicants anonymized examples of the content they will handle. Given the long-term trauma documented in our findings, BPOs can extend mental-health coverage for 6–12 months after employment through negotiated insurance packages. Finally, NDAs can be adapted to allow workers to discuss the emotional toll of their work with trusted family members or mental-health professionals without disclosing platform-specific information.}

\item \revision{As the final clients in the moderation supply chain, \textbf{Social Media Platforms} significantly shape BPO practices. They can strengthen conditions by embedding quarterly, anonymized surveys of moderators’ well-being into their vendor contracts and using these results to enforce minimum standards. Trust and Safety teams should also engage directly with research and worker feedback to ensure platform policies better reflect the realities moderators face.}

 \item \revision{\textbf{Government agencies}, such as the Kenyan Labor Bureau, can strengthen oversight by working with labor associations like the African Content Moderators Union to monitor conditions through regular surveys and workplace inspections. They can enforce timely work-permit processing and require BPOs to meet minimum labor and mental-health standards, while also holding platforms accountable for conditions in their supply chains. Efforts to attract BPO investment should not come at the expense of workers’ safety or rights.}

\item \textbf{Public awareness and recognition} of \revision{the important role of content moderators and the challenges they face.} \revision{Both social media platforms and BPOs should help raise public awareness of content moderation as a profession, such as by posting texts or short videos on websites to explain what this job is about and making transparent the moderation policies.} 
\revision{Our findings show} that social recognition \revision{can} 
make moderators feel valued for what they contribute (i.e., their expertise \cite{abdelkadir2025role}) to their communities, which helps avoid feelings of alienation. \revision{Moreover}, public awareness can also diversify the sources of social support that are crucial for mental wellness by reducing stigma and safety concerns due to misunderstandings about the job. We agree with~\citet{matias_measuring_2024} and the moderators themselves \cite{kleinman_i_2024}: 
akin to first responders in dangerous situations, like firefighters, EMTs, and doctors\revision{, m}oderators are at the forefront of harmful content on the Internet and put their mental health on the line to do this job. Recognizing and respecting moderators' role can help them argue for better working conditions and improved social benefits, which are commonly found for first responders in other fields. \revision{This can start with BPOs allowing moderators to speak with their close ones when they seek some support, by relaxing the Non-Disclosure Agreements.} 
 
\end{enumerate}

\revision{By focusing on actionable recommendations, we expand on recent work within HCI to think about social and policy concerns~\cite{yang_future_2024,lazar2016human}, and classic work on considering the role of whether implications ought to be to design~\cite{baumer_when_2011}. Several promising initiatives currently challenge the existing structure and strive toward meaningful change. For example, the Data Workers Inquiry project is a worker-centered initiative that supports moderators in sharing their experiences through direct inquiries in their workplaces, with guidance from academic researchers \cite{miceli2025methodological}. This innovative approach has gained recognition beyond the academic community. For instance, Fasica Berhane Gebrekidan's inquiry about her personal experience and that of fellow moderators during the Tigray genocide on Facebook was shortlisted for the annual True Story Award alongside established writers and journalists \cite{truestory_content_nodate}. Similarly, moderators in 2023 created the first African Content Moderators Union to collectively demand their rights. Such regional initiatives, which represent a promising starting point for moderators' epistemic authority, collectively represent a promising starting point toward a better future.}


\label{sec:conclusion}
\section{LIMITATIONS AND FUTURE WORK}

Our study may have selection bias \revision{due to participant recruitment} through private groups and snowball sampling. 
While most African moderators are part of union groups \revision{that} were involved in \revision{recruitment}, some voices may be missing\revision{, particularly from those less connected to moderator networks}. Additionally, \revision{despite efforts to recruit moderators from all moderation markets, strict NDAs might prevent moderators from participating in research} 
in the region for fear of retaliation. 

Future work could \revision{include} other actors in the content moderation process, such as \revision{BPO and platform managers}, \revision{counselors or therapists}, and policymakers, to better understand their practices and plans to tackle challenges. More\revision{over}, as with any survey-based research, validity concerns may arise due to reliance on self-reports. Future research could include medical assessments of moderators' long-term health outcomes after moderation work to more comprehensively capture mental health impacts of content moderation.

\section{CONCLUSION}

This paper examines the mental health of commercial content moderators in Africa. Through a mixed-methods approach including surveys (n=134) and contextual inquiry interviews (n=15), we identified the comprehensive and systemic factors that contribute to moderators' severe psychological distress and diminished mental well-being. Our findings reveal that African moderators experience higher psychological distress and lower mental well-being compared to their counterparts in other regions. We discuss how inadequate governmental policies enable BPOs and social media platforms to maintain systemic structures that exacerbate moderators' mental health conditions. Ineffective corporate wellness programs further compound these problems. We found that moderators face long-term impacts that persist well beyond their moderation careers. We argue for the need to think beyond traditional design implications and call for collective action to address these deep-rooted structural and systemic challenges.

\section{ACKNOWLEDGMENT}

\revision{This research is partially supported by a grant from the Internet Society Foundation to the Distributed AI Research Institute (DAIR). We thank our colleagues at the Grouplens lab and DAIR institute for their valuable conversations and feedback, and all research participants who took part in this study. We also thank our anonymous reviewers for their helpful comments in revising this paper.}


\bibliographystyle{ACM-Reference-Format}

\bibliography{added_refs, references_}

\appendix

\revision{
\section{Survey Sample Questions}
\label{surevey_sample_questions_appendix}
Here we provide some of the questions the participants were asked in the survey. After completing the consent procedure, we gathered participants' demographic information, including their names and genders. Some of the other questions asked during the survey include:}

\begin{enumerate}
    \item \revision{What platform did you work for as a moderator?}
    \revision{
    \begin{enumerate}[label=\alph*)]
        \item Meta
        \item TikTok
        \item Other
    \end{enumerate}
    }
    \item \revision{What outsourcing company did you work for? (checkbox)}
    \revision{
    \begin{enumerate}[label=\alph*)]
        \item Sama Source
        \item Teleperformance/Majorel
        \item Concentrix (formerly Webhelp)
        \item Other
    \end{enumerate}
    }

    \item \revision{For how long did you work as a content moderator?}
    \revision{
    \begin{enumerate}[label=\alph*)]
        \item Less than 1 year
        \item 1-2 years
        \item 2-3 years
        \item More than 3 years
    \end{enumerate}
    }

    \item \revision{Are you currently working as a content moderator?}
    \revision{
    \begin{enumerate}[label=\alph*)]
        \item Yes
        \item No
    \end{enumerate}}

    \item \revision{If "No", when did you stop as a content moderator? (Date Picker -- Month, Day, Year)}

    \item \revision{Are you currently employed in other occupations?}
    \revision{
    \begin{enumerate}[label=\alph*)]
        \item Yes
        \item No
    \end{enumerate}}

    \item \revision{How frequently are you/ were you exposed to distressing content? (Choose all that best apply)}
    \revision{
    \begin{enumerate}[label=\alph*)]
        \item Multiple times a day
        \item Few times a day
        \item Multiple times a week
        \item Few times a week
        \item Few times a month
        \item Rarely
    \end{enumerate}}

    \item \revision{Are you currently taking any medications or drugs to help you with your mental well-being?}
    \revision{
    \begin{enumerate}[label=\alph*)]
        \item Yes
        \item No
    \end{enumerate}}

\end{enumerate}

\revision{We then proceeded to the two self-reported measures of psychological distress and mental well-being: the CORE-10 and SWEMWBS, respectively. The CORE-10 questions ask 10 questions where participants select the answer that best describes their feeling in the last week. The questions include: \footnote{For more on CORE-10 -- \url{https://novopsych.com/wp-content/uploads/2024/07/CORE-10-English-NovoPsych.pdf}}}

\begin{itemize}
    \item \revision{I have felt tense, anxious, or nervous.}
    \revision{
    \begin{enumerate}[start=0]
        \item Not at all
        \item Only Occasionally
        \item Sometimes
        \item Often
        \item Most of the time
    \end{enumerate}}
\end{itemize}

\revision{The Short Warwick-Edinburgh Mental Well-being Scale (SWEMWBS) contains 7 questions where participants select the answer that best describes their experience of each over the last two weeks. The questions include: \footnote{For more details on SWEMWBS --\url{https://warwick.ac.uk/services/innovations/wemwbs/how/2025-02-res-area-ww-licenced}}} 

\begin{itemize}
    \item \revision{I’ve been feeling optimistic about the future.}
    \revision{
    \begin{enumerate}
        \item None of the Time
        \item Rarely
        \item Some of the Time
        \item Often
        \item All of the Time
    \end{enumerate}    }
\end{itemize}

\revision{We finally conclude with the factors that primarily affect their mental well-being and what considerations they suggest should be implemented to improve their mental health conditions. The asked questions are:} 

\begin{itemize}
    \item \revision{What do you think are the main factors impacting your mental health as content moderators? \fbox{TEXT AREA}}

\item \revision{In order for you or other content moderators to have a much better well-being, what considerations should be implemented by outsourcing companies and social media platforms? \fbox{TEXT AREA} }   
    
\end{itemize}

\section{\revision{Survey Participant Details}}
\label{survey_participants_detail_appendix}
\revision{Here we provide further details on the survey participants. We provide the context of the former moderators for both platforms (Meta and TikTok), including the time since being a moderator. It shows that the majority of Meta's former moderators stopped working in 2023. This was the year Sama (formerly Sama Source) laid off about 260 moderators in March 2023 after ending its contract with Meta \cite{foxglove_unions_2020}. The majority of the former TikTok moderators stopped working in 2024 and 2025, constituting 43.8\% in each of these two years.}

\begin{table}[h]
\centering
\revision{
\begin{tabular}{lrr}
\toprule
\textbf{Year} & \textbf{Meta (\%)} & \textbf{TikTok (\%)} \\
\midrule
2020 & 2.3 & 3.1 \\
2021 & 2.3 & 0 \\
2022 & 11.6 & 6.3 \\
2023 & 62.8 & 3.1 \\
2024 & 9.3 & 43.8 \\
2025 & 11.6 & 43.8 \\
\bottomrule
\end{tabular}
\caption{\revision{Year distribution of the former moderators for both platforms.}}
}
\end{table}




\end{document}